# Nonempirical definition of the Mendeleev numbers: Organizing the chemical space

Zahed Allahyari[1,*] and Artem R. Oganov [1,†]

[1] Skolkovo Institute of Science and Technology, Skolkovo Innovation Center, 3 Nobel Street, Moscow 143026, Russia

[*]zahed.allahyari@gmail.com, [†]A.Oganov@skoltech.ru

## Abstract

Organizing a chemical space so that elements with similar properties would take neighboring places in a sequence can help to predict new materials. In this paper, we propose a universal method of generating such a one-dimensional sequence of elements, i.e. at arbitrary pressure, which could be used to create a well-structured chemical space of materials and facilitate the prediction of new materials. This work clarifies the physical meaning of Mendeleev numbers, which was earlier tabulated by Pettifor. We compare our proposed sequence of elements with alternative sequences formed by different Mendeleev numbers using the data for hardness, magnetization, enthalpy of formation, and atomization energy. For an unbiased evaluation of the MNs, we compare clustering rates obtained with each system of MNs.

## 1. Introduction

Vast amounts of information about the physical properties and crystal structures of materials have been produced and need to be organized in a clear way to facilitate insight. Even for known materials many properties remain unexplored, and a clear organization of data similar to Mendeleev's Periodic Table would help to estimate these properties *a priori* and uncover those regions of the chemical space that deserve a deeper study.

To solve this challenging problem, it is necessary to construct a coherent chemical space, basically a coordinate system, in which materials with similar properties are closely related and likely to be placed in neighboring regions. This way, prediction of one material would lead to predictions of other materials with similar or perhaps even better properties.

This idea of a chemical space can be explained on a simple example of a set of colored pencils, in which the pencils are put in an order so that the color variation between the adjacent pencils is minimal (Fig. 1). In this example, the pencils represent the elements of the Periodic Table while the colors represent their properties. A combination of two different colors can be considered a binary system in which fractions of colors represent the composition (stoichiometry), while the resulting color shows the properties of the system. A two-dimensional color map, built in such a way, represents a chemical space where binary systems with similar properties are located close to each other, which is the direct result of a suitable one-dimensional arrangement of the elements.

A similar idea of "structure map" was explored in 1984 by Pettifor,[1] who suggested that a well-structured chemical space can be derived by changing the sequence of the elements in the Periodic Table.[1] He proposed a chemical scale that determines the "distance" between the elements on a one-dimensional axis and a Mendeleev number (MN) — an integer showing the position of an element in the sequence.[2] Pettifor claimed that binary compounds with the same structure type occupy the same region in a two-dimensional map plotted using the MNs (the Pettifor map). He evaluated the chemical scale by presenting a map clearly separating 34 different structure types of 574 binary $AB$ compounds (Fig. 2a).[1] Later, Pettifor showed that the MN approach also works for other $A_xB_y$ compounds.[2] Although Pettifor derived the chemical scale and Mendeleev number empirically and based his assessment on only several hundred binary compounds, his study provided a phenomenally successful ordering of the elements confirmed in many later works.[3,4] In this work we denote Pettifor's MN as $MN_P$. We expect that a nonempirical method of finding the MNs would perform even better.

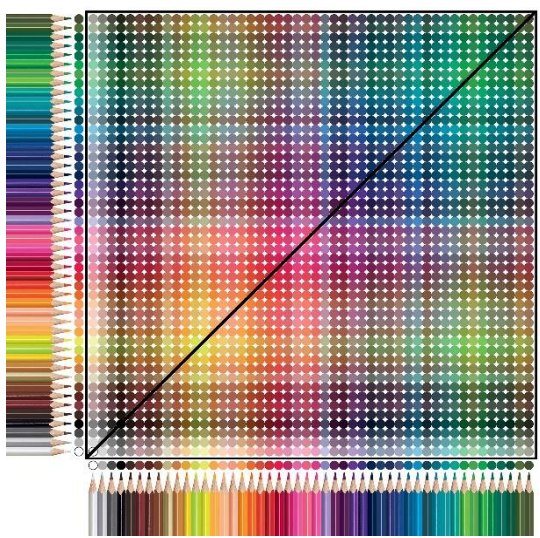

Figure 1. A colored pencil diagram demonstrating the idea of chemical space.

Earlier, in 1929, Goldschmidt tried to find a systematic relationship between the chemical composition and crystal structures of materials. His goal, in particular, was to find how a crystal structure (the geometric arrangement of atoms in a crystal) depends on the chemical composition. The result of his work, known as Goldschmidt's law of crystal chemistry, states that the crystal structure is determined by stoichiometry, atomic size, and polarizability of atoms/ions.[5] In 1955, Goldschmidt's law was modified by Ringwood when he added the electronegativity as another important parameter determining the crystal structure.[6] Based on this premise, we define the chemical scale and MN from these atomic properties.

In 2008, Villars et al. propose a different enumeration of the elements (called periodic number – PN), emphasizing the role of valence electrons.[7] In contrast to the atomic number (AN), PN depends in details on the underlying Periodic Table of the elements.

In 2016, Glawe et al. proposed another sequence of elements (modified MN – in this work we show by $MN_m$) based on their similarity, defining elements A and B to be similar if they crystallize in the same structure type when combined with other elements of the Periodic Table. For example, the alkali metals (Li, Na, K, etc.), forming the rocksalt crystal structure when mixed with Cl, are similar by this definition.[4] Applying this definition and using the available crystal structures in the Inorganic Crystal Structure Database (ICSD),[8] the degrees of similarity of each element with respect to other elements were calculated. Based on these data, the best sequence of elements was optimized using a genetic algorithm, so that similar elements occupy neighboring places in this arrangement.

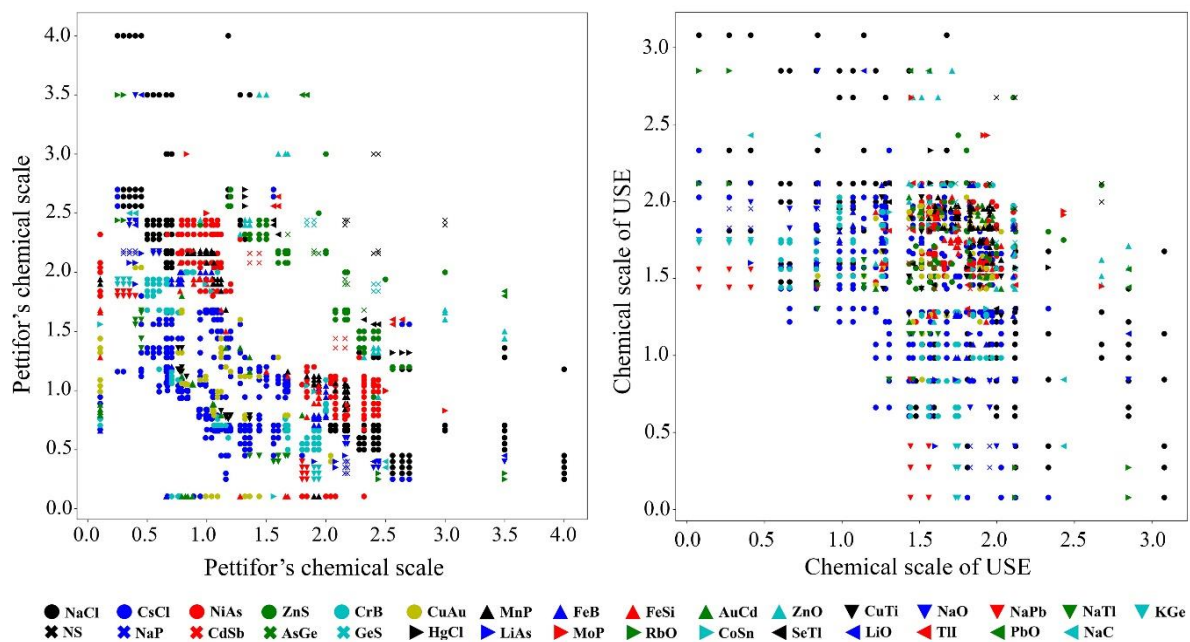

Figure 2. Structure maps of 521 binary *AB* compounds using Pettifor's chemical scale and our redefined chemical scale.

However, defining the MNs with the help of databases has its drawbacks. The first and most important one is that the calculations of the MNs in this case are property-dependent. The quality of the results is lowered because all the structures in the ICSD were taken into account, including theoretical and experimental, stable and metastable at the same time. Also, note that for some elements the data in the ICSD are insufficient.

In this paper, we present a simple, physically meaningful, fully nonempirical universal method of defining the MNs and obtaining the universal sequence of elements (USE). We then compare different MNs using our own theoretical database, which contains about 500,000 crystal structures.

As the chemistry of the elements and materials changes under pressure, so will the MNs. The proposed universal method makes it possible to define the MNs of the elements by their electronegativity and atomic radius at any pressure. In Section 3, we use these properties to compute MNs of a number of elements at high pressures (50 GPa, 200 GPa, and 500 GPa).

## 2. Methods

Unlike Pettifor, who derived his $MN_P$ empirically, we offer a nonempirical (and, therefore, more universal) definition. The most important chemical properties of an atom are the radius $R_a$, electronegativity $\chi$, polarizability $\alpha$, and valence $v$. We disregarded the polarizability in favor of the electronegativity because they are strongly correlated.[9] For simplicity, we also excluded the valence, which is not constant for many elements. Thus, we only consider the electronegativity and atomic radius to define the MNs and obtain the USE (Table 1).

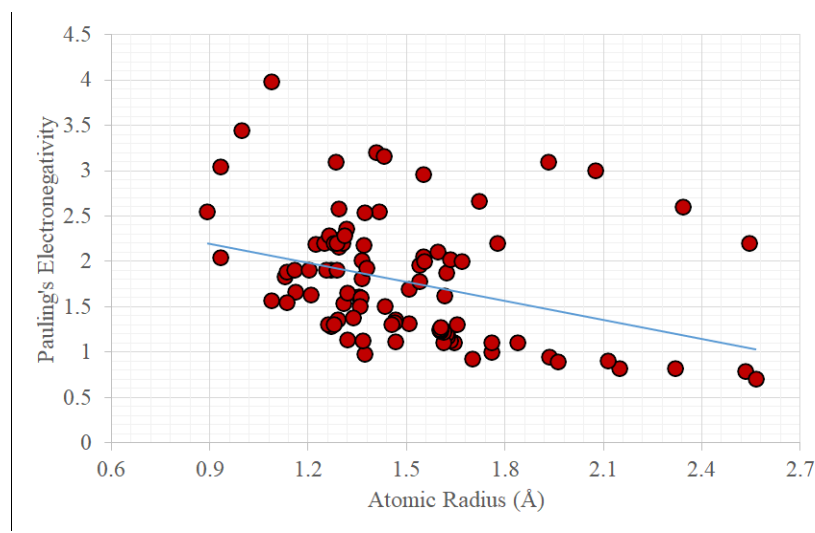

Figure 3. Electronegativities and atomic radii of the elements. The regression line is shown in blue.

We used the Pauling scale for the electronegativity $\chi$.[10] For each element there are many values of atomic radius depending on the bonding type (ionic, covalent, metallic, and van der Waals), oxidation state, and coordination number. The problem is that we need to use values obtained in a consistent way for all elements, and such values were not available. In this work, the atomic radius $R_a$ is defined as half the shortest interatomic distance in the relaxed simple cubic structure of an element. A significant correlation between Pauling's electronegativity $\chi$ and atomic radius $R_a$ (Fig. 3) means that one of them or better some combination of the two can be used as a single parameter approximately characterizing the chemistry of an element. To find an approximate combination of these two parameters into one, the regression line in the space of $\chi$ and $R_a$ was computed and all the elements were projected onto it (Fig. 3). The zero value on this scale was assigned to the projection of the first element (the one having a large atomic radius and low electronegativity) onto the regression line, while the coordinates of other elements on the line were defined as the distance of their projections from zero – these are defined as the chemical scale. The Mendeleev number, USE, was defined as the sequential number of the projected element on the regression line (see Table 2).

Table 1. Electronegativities and atomic radii of the elements used for obtaining the universal sequence of elements (USE).

| Element | Atomic radius $R_a$ (Å) | Pauling Electronegativity ($\chi$) | Element | Atomic radius $R_a$ (Å) | Pauling Electronegativity ($\chi$) |
|---|---|---|---|---|---|
| H | 0.727 | 2.2 | In | 1.541 | 1.78 |
| He | 1.286 | 3.1 | Sn | 1.541 | 1.96 |
| Li | 1.374 | 0.98 | Sb | 1.553 | 2.05 |
| Be | 1.090 | 1.57 | Te | 1.596 | 2.1 |
| B | 0.933 | 2.04 | I | 1.721 | 2.66 |
| C | 0.891 | 2.55 | Xe | 2.344 | 2.6 |
| N | 0.932 | 3.04 | Cs | 2.535 | 0.79 |
| O | 0.997 | 3.44 | Ba | 1.962 | 0.89 |
| F | 1.089 | 3.98 | La | 1.647 | 1.1 |
| Ne | 1.409 | 3.2 | Ce | 1.467 | 1.12 |
| Na | 1.701 | 0.93 | Pr | 1.367 | 1.13 |
| Mg | 1.508 | 1.31 | Nd | 1.320 | 1.14 |
| Al | 1.355 | 1.61 | Pm | 1.635 | 1.13 |
| Si | 1.269 | 1.9 | Sm | 1.626 | 1.17 |
| P | 1.223 | 2.19 | Eu | 1.620 | 1.2 |
| S | 1.293 | 2.58 | Gd | 1.623 | 1.2 |
| Cl | 1.431 | 3.16 | Tb | 1.613 | 1.1 |
| Ar | 1.933 | 3.1 | Dy | 1.613 | 1.22 |
| K | 2.151 | 0.82 | Ho | 1.604 | 1.23 |
| Ca | 1.761 | 1 | Er | 1.602 | 1.24 |
| Sc | 1.466 | 1.36 | Tm | 1.602 | 1.25 |
| Ti | 1.308 | 1.54 | Yb | 1.759 | 1.1 |
| V | 1.209 | 1.63 | Lu | 1.605 | 1.27 |
| Cr | 1.162 | 1.66 | Hf | 1.454 | 1.3 |
| Mn | 1.136 | 1.55 | Ta | 1.358 | 1.5 |
| Fe | 1.131 | 1.83 | W | 1.316 | 2.36 |
| Co | 1.137 | 1.88 | Re | 1.287 | 1.9 |
| Ni | 1.160 | 1.91 | Os | 1.278 | 2.2 |
| Cu | 1.203 | 1.9 | Ir | 1.288 | 2.2 |
| Zn | 1.320 | 1.65 | Pt | 1.311 | 2.28 |
| Ga | 1.365 | 1.81 | Au | 1.374 | 2.54 |
| Ge | 1.365 | 2.01 | Hg | 1.556 | 2 |
| As | 1.369 | 2.18 | Tl | 1.617 | 1.62 |
| Se | 1.418 | 2.55 | Pb | 1.622 | 1.87 |
| Br | 1.551 | 2.96 | Bi | 1.635 | 2.02 |
| Kr | 2.077 | 3 | Po | 1.670 | 2 |
| Rb | 2.319 | 0.82 | At | 1.777 | 2.2 |
| Sr | 1.935 | 0.95 | Rn | 2.544 | 2.2 |
| Y | 1.625 | 1.22 | Fr | 2.567 | 0.7 |
| Zr | 1.463 | 1.33 | Ra | 2.114 | 0.9 |
| Nb | 1.362 | 1.6 | Ac | 1.838 | 1.1 |
| Mo | 1.294 | 2.16 | Th | 1.655 | 1.3 |
| Tc | 1.257 | 1.9 | Pa | 1.436 | 1.5 |
| Ru | 1.249 | 2.2 | U | 1.339 | 1.38 |
| Rh | 1.264 | 2.28 | Np | 1.291 | 1.36 |
| Pd | 1.306 | 2.2 | Pu | 1.271 | 1.28 |
| Ag | 1.379 | 1.93 | Am | 1.261 | 1.3 |
| Cd | 1.509 | 1.69 | Cm | 1.279 | 1.3 |

## 3. Results and Discussion

In a well-ordered sequence of elements, the atoms with similar properties are close to each other. Therefore, in the two-dimensional chemical space based on such sequence, the properties of neighboring binary systems should exhibit a close relation. On this premise, we evaluate different MNs: atomic number (AN), Villars' periodic number [7] (PN), Pettifor's Mendeleev number [2] ($MN_P$), modified Mendeleev number [4] ($MN_m$), and Mendeleev number in this work, the universal sequence of elements, (USE). These MNs are shown in Table 2.

To examine different MNs, a database containing about 500,000 theoretical and experimental crystal structures of unary and binary compounds was compiled. These structures were relaxed using density functional theory within the generalized gradient approximation (DFT-GGA) and the database was set up so as to contain neither duplicates nor very unstable structures (whose energy is more than 0.5 eV/atom above the convex hull). Some crystal structures in the database were imported from other online databases, such as ICSD [8] and COD,[11] while the majority came from the previous calculations based on the evolutionary algorithm USPEX.[12–14]

The database contains the crystal structure information for 1,591 binary and 80 unary systems – excluding Ar, Xe, Ce, Nd, Pm, Sm, Eu, Gd, Tb, Dy, Ho, Er, Tm, Yb, Lu, Rn, Bk, Cf, Es, Fm, Md, No, Lr, Rf, and Db. Of these, only 446 systems have the magnetic information that are obtained in several multi-objective evolutionary searches for low-energy and highly magnetized phases, as implemented in the USPEX algorithm.[15] The hardness of all crystal structures in this database was computed using the Lyakhov-Oganov model.[16] The database is fully consistent because all crystal structures were relaxed and their energies computed in the same settings using the density functional theory with the projector-augmented wave method (PAW) and PBE [17] functional as implemented in the VASP code.[18,19] To compare the performance of different MNs for binary systems, the 2D maps of various properties were plotted, among them the hardness (representing the mechanical properties), magnetization (electronic properties), enthalpy of formation, and atomization energy (thermodynamic and chemical properties).

For hardness and magnetization, the representative structure of each binary system is a structure with the energy less than 0.1 eV/atom above the convex hull, having the highest hardness or magnetization, respectively. For generating the chemical spaces of the enthalpy of formation and atomization energy, the representative structure of each binary system is a structure with the lowest enthalpy of formation or lowest atomization energy, respectively. In all cases, no restrictions on stoichiometries of studied structures were imposed. The generated chemical spaces of hardness, magnetization, enthalpy of formation, and atomization energy using different MNs are shown in Fig. 4,5,6, and 7, respectively.

Table 2. The universal sequence of elements (USE), coordinates of the elements on the regression line – chemical scale (CS), atomic number (AN), periodic number [7] (PN), Pettifor's Mendeleev number [2] ($MN_P$), modified MN [4] ($MN_m$).

| # | USE | CS | AN | PN | $MN_P$ | $MN_m$ | # | USE | CS | AN | PN | $MN_P$ | $MN_m$ |
|---|---|---|---|---|---|---|---|---|---|---|---|---|---|
| 1 | Fr | 0 | H | Li | He | He | 51 | Bi | 1.517 | Sb | Re | V | V |
| 2 | Cs | 0.077 | He | Na | Ne | Ne | 52 | Sn | 1.560 | Te | Fe | W | Cr |
| 3 | Rb | 0.272 | Li | K | Ar | Ar | 53 | Zn | 1.566 | I | Ru | Mo | Mo |
| 4 | K | 0.411 | Be | Rb | Kr | Kr | 54 | Hg | 1.571 | Xe | Os | Cr | W |
| 5 | Ra | 0.486 | B | Cs | Xe | Xe | 55 | Te | 1.594 | Cs | Co | Tc | Re |
| 6 | Ba | 0.606 | C | Fr | Rn | Rn | 56 | Sb | 1.601 | Ba | Rh | Re | Tc |
| 7 | Sr | 0.662 | N | Ca | Fr | Fr | 57 | Ga | 1.620 | La | Ir | Mn | Os |
| 8 | Ac | 0.827 | O | Sr | Cs | Cs | 58 | V | 1.646 | Ce | Ni | Fe | Ru |
| 9 | Ca | 0.834 | F | Ba | Rb | Rb | 59 | Mn | 1.661 | Pr | Pd | Os | Ir |
| 10 | Na | 0.843 | Ne | Ra | K | K | 60 | Ag | 1.676 | Nd | Pt | Ru | Rh |
| 11 | Rn | 0.871 | Na | Sc | Na | Na | 61 | Cr | 1.702 | Pm | Cu | Co | Pt |
| 12 | Yb | 0.892 | Mg | Y | Li | Li | 62 | Be | 1.710 | Sm | Ag | Ir | Pd |
| 13 | La | 0.984 | Al | La | Ra | Ra | 63 | Kr | 1.710 | Eu | Au | Rh | Au |
| 14 | Pm | 1.011 | Si | Ac | Ba | Ba | 64 | Ge | 1.733 | Gd | Be | Ni | Ag |
| 15 | Tb | 1.012 | P | Ce | Sr | Sr | 65 | Re | 1.735 | Tb | Mg | Pt | Cu |
| 16 | Sm | 1.041 | S | Th | Ca | Ca | 66 | Si | 1.750 | Dy | Zn | Pd | Ni |
| 17 | Gd | 1.061 | Cl | Pr | Yb | Eu | 67 | Tc | 1.760 | Ho | Cd | Au | Co |
| 18 | Eu | 1.063 | Ar | Pa | Eu | Yb | 68 | Cu | 1.804 | Er | Hg | Ag | Fe |
| 19 | Y | 1.071 | K | Nd | Y | Lu | 69 | I | 1.810 | Tm | B | Cu | Mn |
| 20 | Dy | 1.081 | Ca | U | Sc | Tm | 70 | Fe | 1.824 | Yb | Al | Mg | Mg |
| 21 | Th | 1.091 | Sc | Pm | Lu | Y | 71 | As | 1.827 | Lu | Ga | Hg | Zn |
| 22 | Ho | 1.094 | Ti | Np | Tm | Er | 72 | Ni | 1.845 | Hf | In | Cd | Cd |
| 23 | Er | 1.101 | V | Sm | Er | Ho | 73 | Co | 1.847 | Ta | Tl | Zn | Hg |
| 24 | Tm | 1.107 | Cr | Pu | Ho | Dy | 74 | Mo | 1.877 | W | C | Be | Be |
| 25 | Lu | 1.116 | Mn | Eu | Dy | Tb | 75 | Ar | 1.885 | Re | Si | Tl | Al |
| 26 | Li | 1.141 | Fe | Am | Tb | Gd | 76 | Pd | 1.890 | Os | Ge | In | Ga |
| 27 | Ce | 1.144 | Co | Gd | Gd | Sm | 77 | Ir | 1.905 | Ir | Sn | Al | In |
| 28 | Mg | 1.218 | Ni | Cm | Sm | Pm | 78 | Os | 1.913 | Pt | Pb | Ga | Tl |
| 29 | Pr | 1.232 | Cu | Tb | Pm | Nd | 79 | Pt | 1.931 | Au | N | Pb | Pb |
| 30 | Hf | 1.257 | Zn | Bk | Nd | Pr | 80 | Ru | 1.937 | Hg | P | Sn | Sn |
| 31 | Xe | 1.263 | Ga | Dy | Pr | Ce | 81 | P | 1.953 | Tl | As | Ge | Ge |
| 32 | Zr | 1.266 | Ge | Cf | Ce | La | 82 | Rh | 1.970 | Pb | Sb | Si | Si |
| 33 | Nd | 1.276 | As | Ho | La | Ac | 83 | W | 1.973 | Bi | Bi | B | B |
| 34 | Sc | 1.281 | Se | Es | Fm | Th | 84 | Se | 1.997 | Po | O | Bi | C |
| 35 | Tl | 1.304 | Br | Er | Es | Pa | 85 | Au | 2.027 | At | S | Sb | N |
| 36 | Pa | 1.385 | Kr | Fm | Cf | U | 86 | B | 2.106 | Rn | Se | As | P |
| 37 | Pu | 1.396 | Rb | Tm | Bk | Np | 87 | S | 2.116 | Fr | Te | P | As |
| 38 | U | 1.397 | Sr | Yb | Cm | Pu | 88 | Br | 2.120 | Ra | Po | Po | Sb |
| 39 | Cm | 1.401 | Y | Lu | Am | Am | 89 | Cl | 2.332 | Ac | H | Te | Bi |
| 40 | Am | 1.416 | Zr | Ti | Pu | Cm | 90 | H | 2.366 | Th | F | Se | Po |
| 41 | Np | 1.425 | Nb | Zr | Np | Bk | 91 | Ne | 2.373 | Pa | Cl | S | Te |
| 42 | Cd | 1.433 | Mo | Hf | U | Cf | 92 | He | 2.418 | U | Br | C | Se |
| 43 | Pb | 1.442 | Tc | V | Pa | Es | 93 | C | 2.430 | Np | I | At | S |
| 44 | Ta | 1.449 | Ru | Nb | Th | Fm | 94 | N | 2.675 | Pu | At | I | O |
| 45 | In | 1.458 | Rh | Ta | Ac | Sc | 95 | O | 2.849 | Am | He | Br | At |
| 46 | Po | 1.477 | Pd | Cr | Zr | Zr | 96 | F | 3.080 | Cm | Ne | Cl | I |
| 47 | At | 1.502 | Ag | Mo | Hf | Hf | 97 |  |  | Bk | Ar | N | Br |
| 48 | Nb | 1.503 | Cd | W | Ti | Ti | 98 |  |  | Cf | Kr | O | Cl |
| 49 | Ti | 1.513 | In | Mn | Nb | Ta | 99 |  |  | Es | Xe | F | F |
| 50 | Al | 1.514 | Sn | Tc | Ta | Nb | 100 |  |  | Fm | Rn | H | H |

### 3.1. Evaluation of MNs

In a correctly defined chemical space, closely located materials should have similar properties. The most promising materials will then be clustered in one or few "islands" in this space. To predict new materials, it could be sufficient to explore these islands instead of the entire chemical space. The fewer these islands are, the easier it would be to locate and explore them for promising materials. A chemical space containing many small islands is less amenable for the prediction of materials than the one with fewer big islands. Therefore, for evaluating each chemical space, it is useful to find these islands and calculate the number of (similar) materials they cover.

For doing this, we used the idea of the clustering algorithm proposed by Rodriguez and Laio [20] and applied it to clustering regions of the chemical space on the basis of their similarity. In this simple method, each cluster is defined by a cluster center and a number of similar data points around it. For finding the cluster centers, two quantities are to be calculated for each data point $i$: its local density $\rho_i$, and its distance $\delta_i$ from the nearest point with a higher density. In the original method, $\rho_i$ is equal to the number of points that are closer than $d_c$ to the point $i$ (we call these points: local neighbors), where $d_c$ is a cutoff radius. Also, $\delta_i$ for the point with the highest density is equal to its distance from the furthest data point. This way, the cluster centers are those points with high value of both $\rho$ and $\delta$. Clearly, the point with the highest density $\rho_i$, is always a cluster center.

In our modified method, we only consider the point with the highest density as a cluster center, and therefore, there is no need for calculation of $\delta_i$. Then, we remove the cluster center and all its local neighbors from the dataset, we calculate $\rho_i$ again for the remaining data points, and find a new cluster center. We continue this loop until all the data points are assigned to a cluster. The points with zero local density $\rho_i$ are isolated points. In our method, $\rho_i$ is equal to the number of points that are closer than $d_c$ to the point $i$, and their property difference to the point $i$ is less than $d_p$, where $d_p$ is a property difference cutoff. We need to clarify that data points closer than $d_c$ to a local neighbor – neighbors of the local neighbors – with property difference less than $d_p$ from the cluster center, are also included in the cluster and considered as local neighbors of the cluster center, but these points are not included in calculation of local density $\rho_i$ in the first place.

The number of clusters (i.e. islands) that cover all binary systems in the chemical spaces of the MNs, is a good quantitative evaluation of these MNs. The lower the number of clusters, the better-clustered the chemical space. However, as cutoff values, i.e. $d_c$ and $d_p$, are increased, the number of clusters decreases (see Fig. 8).

For finding the cluster centers, the constant cutoff radius $d_c$ equal to 5 blocks was used – clusters expand by including neighbors of the local neighbors as mentioned above, and $d_c$ is only used to bound the neighborhood area of each system. To see how number of clusters in different MNs changes with respect to $d_p$, we let this value changes as shown in Fig. 8 and Fig. 9.

Another quantitative evaluation of the MNs is the number of systems that are covered by clusters. For this purpose, we define an imaginary "ideal MN" ($MN_{ideal}$) for each property, which clusters all the materials in a minimum number of clusters ($N_{min}$) in the target chemical space. $N_{min}$ can be easily calculated by having the property range of distributed systems in a chemical space (as shown in the color bar of Fig. 4,5,6, and 7) and $d_p$ (maximum property difference between a cluster member and the cluster center) – the range of this value with regard to the change of the $d_p$ is shown in Table 3 for $MN_{ideal}$. Therefore, our second evaluation criterion is the fraction of all systems that are covered by the first (biggest) $N_{min}$ clusters – the results of this evaluation are shown in Fig. 9. These two evaluations provide an insight into the clustering rate of different MNs.

As mentioned earlier, only 1591 binary and 80 unary systems are studied in our database which is about a half of the total binary and unary systems that can be created from the combination of 80 elements – totally 3240 systems can be created. Of these, hardness, enthalpy of formation, and atomization energy are presented for almost all the studied systems (about 50% of total systems) while magnetization was computed only in 446 systems (about 14% of total systems). The amount of missing information can influence the correct clustering of the chemical space – for example, when a cluster cannot expand because of the lack of data points around it, and not because of the existence of dissimilar systems around it. To solve this problem, we assigned a value to the property of each missing system by cubic interpolation of its neighbors' property in the scale of each MN. Then the property of the missing system is calculated as the average of its values in different MN scales – in the spirit of the *committee voting* approach. We evaluated our *committee voting* approach, by removing materials with explicitly calculated properties in our database, and predicting their properties using *committee voting*. On average, the error (difference between the predicted and calculated values) of the predicted values are: 3.24 GPa for hardness, 0.014 $\mu_\beta/Å^3$ for magnetization, 0.175 eV/atom for the enthalpy of formation, and 0.48 eV/atom for atomization energy – between 3.5% (for enthalpy of formation) to 7% (for magnetization) of the property ranges.

In the following, we discuss different MNs by calculating their clustering rate and visualizing their 2D maps (Pettifor maps) of the hardness, magnetization, enthalpy of formation, and atomization energy.

### 3.2. Hardness

The hardest structure with the energy less than 0.1 eV/atom above the convex hull is the representative structure of a binary system. To get a more accurate map of hardness, the hardnesses of these representative structures were calculated using the Mazhnik-Oganov model [21] of hardness. Then the hardness of the missed systems is calculated using the committee voting method (see Fig. 4). In clustering the hardness maps, we only include materials harder than 5 GPa because the majority of materials are soft (with hardness less than 5 GPa) which is not interesting for us and reduces the difference of the clustering rates of different MNs.

Hardest materials are usually compounds of carbon, boron, and nitrogen with each other or with other elements. When these three elements sit in neighboring places (i.e. in AN and $MN_m$), a number of big islands are produced depending on the arrangement of other elements. Despite that, if other similar elements are placed far from each other, they form several distant islands that are not clustered together (see Fig. 4a). Table 3 shows the number of clusters to cover all binary systems (harder than 5 GPa) in the hardness maps of the MNs. These results are shown in more details in Fig. 8. The maximum number of clusters (islands), in small $d_p$, is found for AN which was expected due to the splotchy hardness map it produced. In the $N_{min}$ biggest clusters, AN covers fewer binary systems than all other MNs in different range of $d_p$ (see Fig. 9). The highest clustering rate is calculated for USE that clusters regions of hardness map in lower number of clusters, than other MNs, in whole range of $d_p$ (Fig. 8). Fig.9 shows that USE covers 78% to 96% (for different $d_p$) of all materials harder than 5 GPa in its biggest $N_{min}$ clusters. Better clustering of materials with similar hardness by USE was expected even by visualizing the produced Pettifor maps of hardness. USE has significantly reduced the size of the regions containing materials harder than 15 GPa – exploring about a quarter of the chemical space is enough to predict almost all the hard materials – that also places soft materials in each other's vicinity. Reducing the size of promising regions of the chemical space is important, especially, when doing an automatic and systematic search for materials with optimal properties.[22]

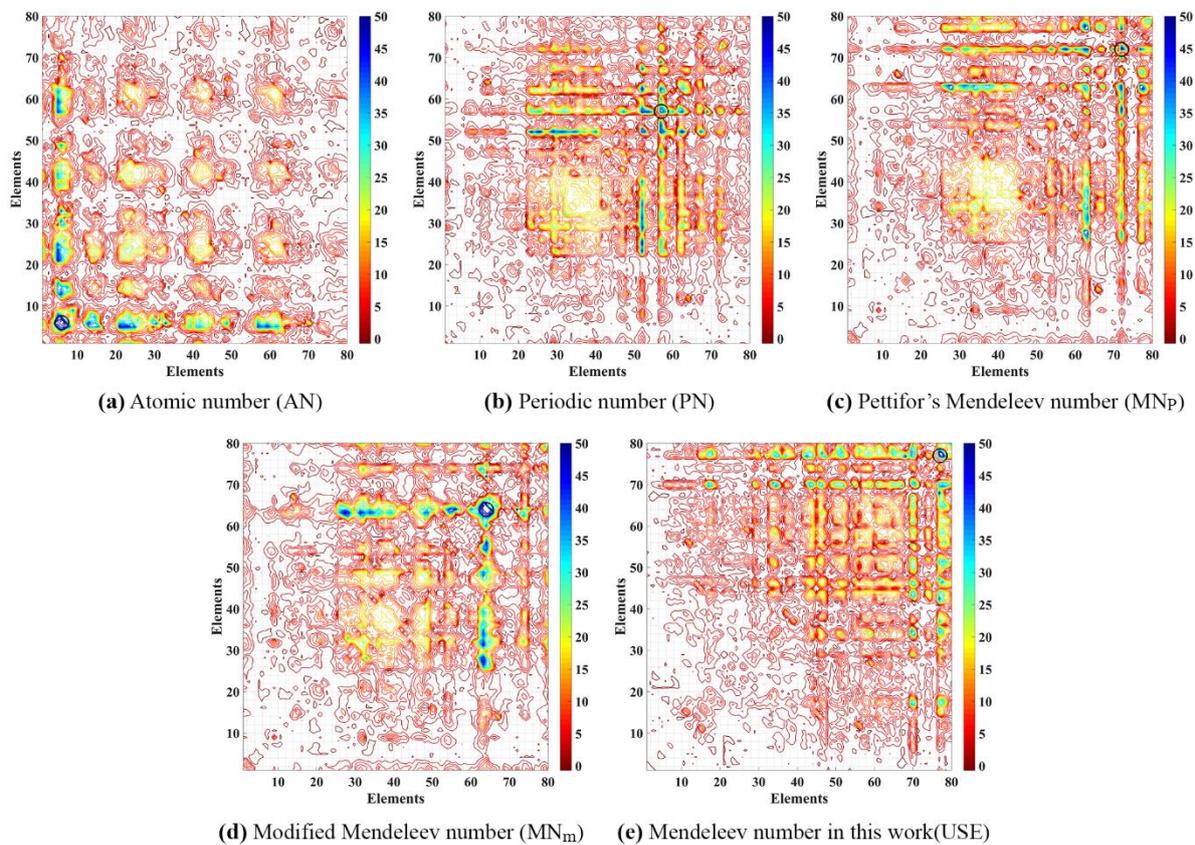

(a) Atomic number (AN)  (b) Periodic number (PN)  (c) Pettifor's Mendeleev number ($MN_P$)

(d) Modified Mendeleev number ($MN_m$)  (e) Mendeleev number in this work(USE)

Figure 4. 2D maps of the hardness (GPa) obtained using Mazhnik-Oganov's model [21] of hardness of binary systems, plotted in various MNs. The representative for each binary system is the phase with the highest hardness in our database. The material with the highest hardness is shown by black hollow circle.

Table 3. Clustering rate based on: (a) the number of clusters for different MNs in comparison to the minimum number of clusters, $N_{min}$, in a imaginary ideal MN ($MN_{ideal}$), and (b) fraction of binary systems that are covered by the first (biggest) $N_{min}$ clusters in different MNs. The clustering rates are calculated based on the change of the $d_p$.

|  | | (a) Number of clusters to cover all binary systems | | | | | (b) Fraction of binary systems that are covered by the first $N_{min}$ clusters | | | | |
| --- | --- | --- | --- | --- | --- | --- | --- | --- | --- | --- | --- |
|  | $d_p$ | $N_{min}$ | AN | PN | $MN_P$ | $MN_m$ | USE | AN | PN | $MN_P$ | $MN_m$ | USE |
| Hardness | 1.5 | 15 | 98 | 87 | 74 | 80 | 65 | 0.59 | 0.68 | 0.71 | 0.7 | 0.78 |
|  | 2.5 | 9 | 62 | 53 | 50 | 55 | 44 | 0.72 | 0.77 | 0.8 | 0.79 | 0.83 |
|  | 3.5 | 7 | 43 | 39 | 39 | 51 | 27 | 0.79 | 0.83 | 0.85 | 0.79 | 0.89 |
|  | 4.5 | 5 | 32 | 37 | 28 | 43 | 24 | 0.8 | 0.84 | 0.88 | 0.82 | 0.9 |
|  | 5.5 | 5 | 29 | 33 | 26 | 34 | 23 | 0.83 | 0.87 | 0.91 | 0.85 | 0.91 |
|  | 6.5 | 4 | 21 | 35 | 20 | 26 | 14 | 0.89 | 0.87 | 0.93 | 0.89 | 0.96 |
| Magnetization | 0.005 | 18 | 70 | 48 | 60 | 63 | 64 | 0.83 | 0.88 | 0.84 | 0.85 | 0.84 |
|  | 0.01 | 9 | 36 | 30 | 30 | 39 | 35 | 0.89 | 0.91 | 0.92 | 0.9 | 0.92 |
|  | 0.02 | 5 | 20 | 15 | 12 | 15 | 14 | 0.94 | 0.96 | 0.97 | 0.97 | 0.97 |
|  | 0.03 | 3 | 12 | 10 | 11 | 8 | 9 | 0.95 | 0.98 | 0.97 | 0.98 | 0.97 |
|  | 0.04 | 3 | 11 | 8 | 10 | 8 | 9 | 0.95 | 0.99 | 0.98 | 0.98 | 0.98 |
|  | 0.05 | 2 | 11 | 5 | 8 | 6 | 6 | 0.96 | 0.99 | 0.99 | 0.98 | 0.99 |
| Enthalpy of formation | 0.05 | 50 | 240 | 197 | 196 | 182 | 193 | 0.75 | 0.78 | 0.79 | 0.79 | 0.79 |
|  | 0.1 | 25 | 136 | 95 | 95 | 87 | 100 | 0.83 | 0.9 | 0.9 | 0.89 | 0.88 |
|  | 0.2 | 13 | 76 | 54 | 47 | 59 | 60 | 0.88 | 0.93 | 0.95 | 0.93 | 0.93 |
|  | 0.3 | 9 | 59 | 37 | 39 | 39 | 46 | 0.93 | 0.96 | 0.95 | 0.95 | 0.95 |
|  | 0.4 | 7 | 35 | 28 | 24 | 31 | 29 | 0.94 | 0.97 | 0.98 | 0.96 | 0.97 |
|  | 0.5 | 5 | 24 | 21 | 21 | 21 | 24 | 0.96 | 0.97 | 0.96 | 0.98 | 0.96 |
| Atomization energy | 0.1 | 44 | 439 | 325 | 321 | 350 | 440 | 0.33 | 0.46 | 0.51 | 0.44 | 0.35 |
|  | 0.2 | 22 | 194 | 141 | 158 | 164 | 190 | 0.53 | 0.76 | 0.71 | 0.63 | 0.64 |
|  | 0.3 | 15 | 130 | 96 | 80 | 89 | 94 | 0.69 | 0.8 | 0.79 | 0.77 | 0.84 |
|  | 0.4 | 11 | 110 | 54 | 52 | 63 | 84 | 0.71 | 0.86 | 0.87 | 0.8 | 0.86 |
|  | 0.5 | 9 | 76 | 49 | 50 | 57 | 70 | 0.76 | 0.89 | 0.87 | 0.8 | 0.88 |
|  | 0.6 | 8 | 65 | 37 | 33 | 47 | 47 | 0.82 | 0.9 | 0.89 | 0.88 | 0.89 |

### 3.3. Magnetization

As mentioned before, our database contains magnetic information for only 14% of the total binary systems and the magnetic information is assigned to the majority of the systems (about 86%) using committee voting method. This, obviously, increases the clustering rate for all the MNs. The range of magnetization is from zero for nonmagnetic and antiferromagnetic materials to 0.198 $\mu_\beta/Å^3$ corresponding to the magnetization of iron. Among all elements that we included, and their compounds, Fe has the highest magnetization. This result is correct. In reality, Gd has a slightly higher magnetization, but lanthanoids were not included for technical reasons (problems with available pseudopotentials, and with convergence). For evaluating magnetization maps of different MNs, we disregarded materials with magnetization less than 0.02 $\mu_\beta/Å^3$ (see Table 3) – this helps us to better distinguish performance of different MNs.

Looking at Fig. 5, at the first glance, it seems that AN provides a slightly better map with clear separation of materials with similar magnetization. Although that might be true for promising regions, a closer look to Fig. 5a shows that AN clusters other regions of the chemical space, with lower magnetization, less efficiently by following its periodic pattern – see Table

3, Fig. 8, and Fig. 9. The clustering rate for all the MNs are very high as could be expected – small number of clusters that quickly approaches the $N_{min}$ (minimum number of clusters that is required by an ideal MN), and high coverage of binary systems (from 83% for small $d_p$, to 99% for bigger $d_p$) in the first $N_{min}$ clusters.

In Fig. 5, the main two islands of materials with high magnetization, correspond to the compounds of some transition metals such as Fe, Co, Ni, and some actinoids such as Pu (lanthanoids also form highly magnetic phases, but as we mentioned above, were excluded for technical reasons) – this can be clearly seen in the magnetization map of AN (Fig. 5a).

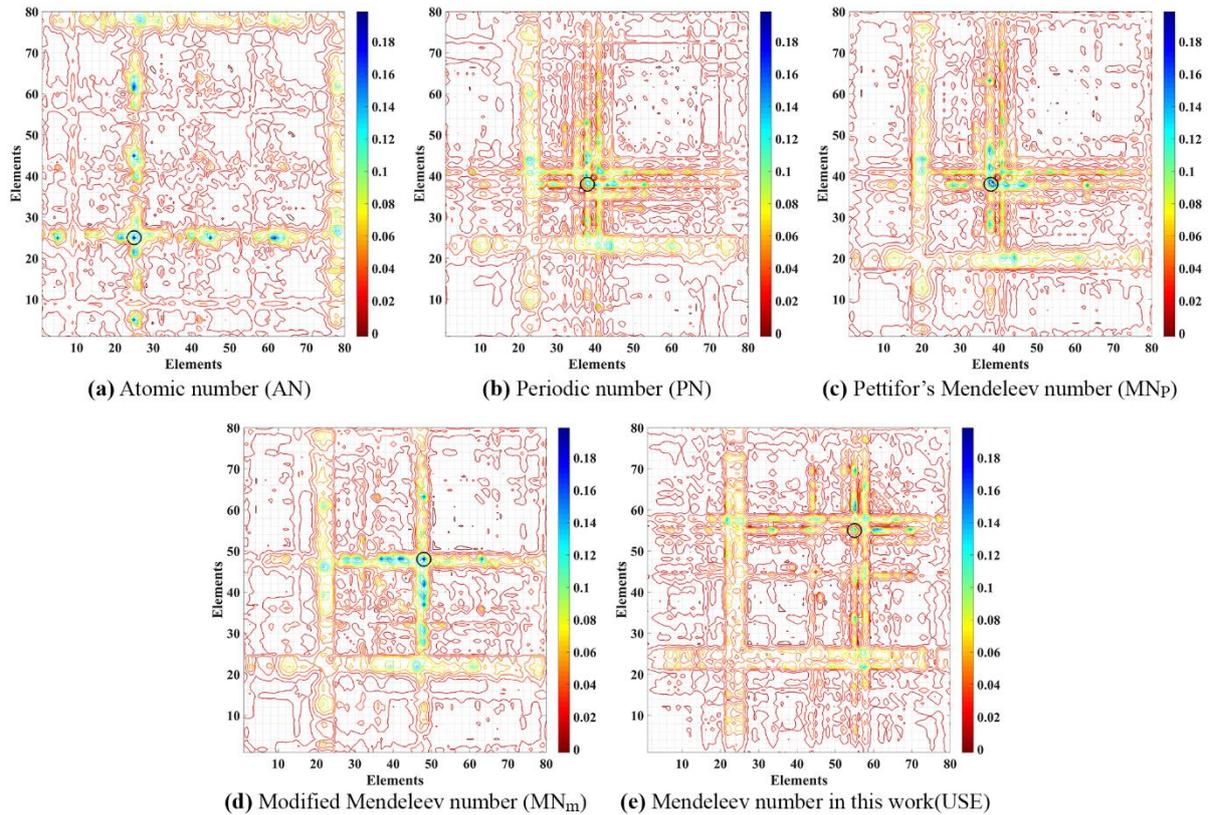

Figure 5. 2D maps of magnetization (in the unit of $\mu_B \cdot Å^{-3}$) of binary systems, plotted in various MNs. The representative for each binary system is the phase with the highest magnetization in our database. The material with the highest magnetization is shown by black hollow circle.

### 3.4. Enthalpy of formation

Pettifor maps of the enthalpy of formation produced by different MNs are shown in Fig. 6. The plots were made taking in each binary system the $A_xB_y$ compound with the lowest enthalpy of formation in the database. Unlike hardness and magnetization map, in the maps of the enthalpy of formation, binary systems with lower values of the enthalpy of formation are more favorable (depicted with red color).

Looking at Fig. 6, one can see that PN, $MN_P$, $MN_m$, and USE, have produced similar maps of the enthalpy of formation. In all these maps, promising materials (with more exothermic chemical reactions – shown in orange and red) are gathered in a small region, right bottom – left top, of the map. This means that very dissimilar elements often form stable compounds. The lowest enthalpy of formation was found for $ThF_4$ (−4.11 eV/atom), followed by $AcF_3$

(−4.09 eV/atom), CaF$_2$ (−3.92 eV/atom) and ZrF$_4$ (−3.62 eV/atom). Other notable values include Th$_4$O$_7$ (−3.61 eV/atom), Y$_2$O$_3$ (−3.48 eV/atom), Al$_2$Ta (−3.18 eV/atom), Al$_2$O$_3$ (−2.95 eV/atom), CaO (−2.95 eV/atom), SiO$_2$ (−2.79 eV/atom), Al$_5$Ge$_2$ (−2.44 eV/atom). Note that fluorides and oxides are the most exothermic compounds, which is easy to understand, since F and O have the highest electronegativities. Materials with a higher enthalpy of formation (shown by dark and light blue) occupy a wide region in the center of the maps. Materials that are shown in yellow color (with an enthalpy of formation between −2 and −2.5 eV/atom) can be found mostly around the promising regions (shown in red). Fig. 8 shows that among these MNs, MN$_m$ provides a slightly better map, while the performance of all MNs (except for AN) are similarly good. Fig. 6 suggests that USE performs better in clustering promising regions (shown in yellow, orange and red) by condensing them in a smaller area. On the other hand, AN produced a periodic map which is inefficient for clustering compounds with similar enthalpies of formation. As expected, our clustering evaluations show that AN clusters regions of the chemical space less efficiently than other MNs (see Fig. 8, and Fig. 9).

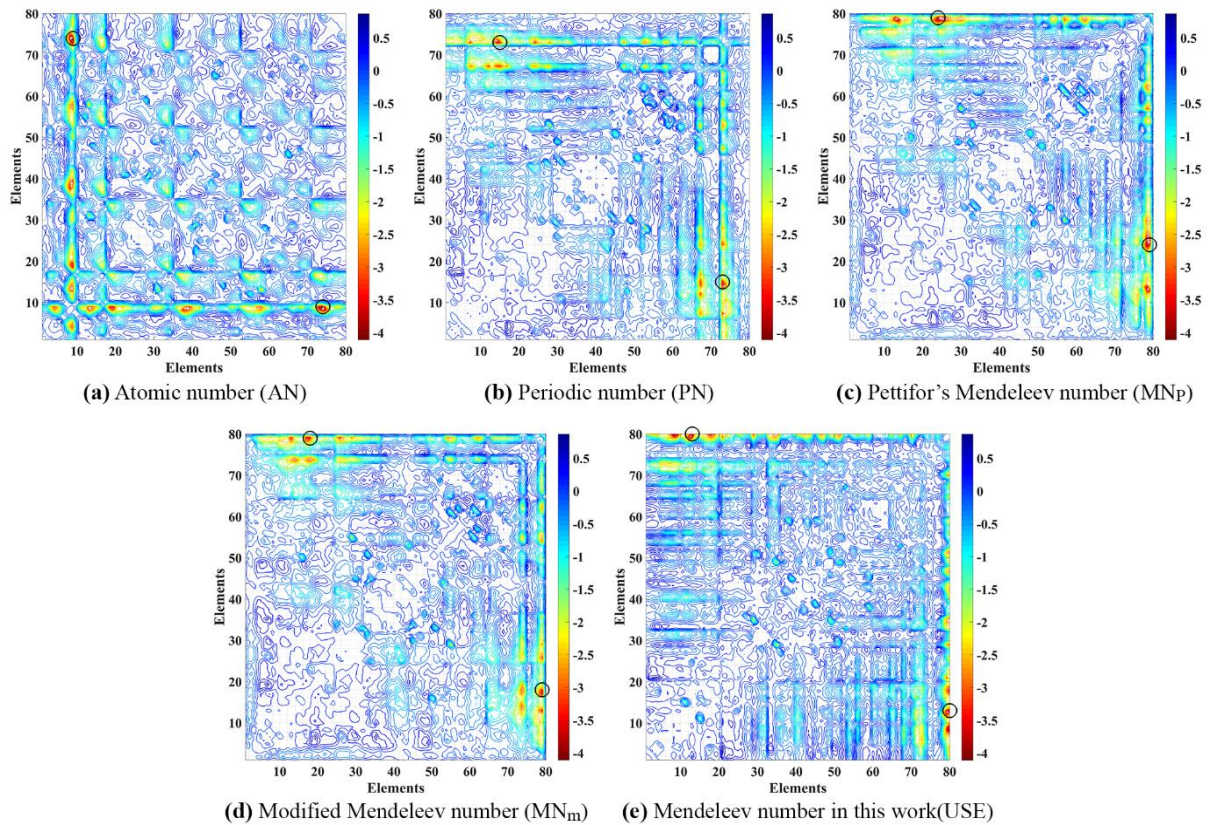

Figure 6. 2D maps of the enthalpy of formation (eV/atom) of binary systems, plotted in various MNs. The representative for each binary system is a structure with the lowest enthalpy of formation in our database. The material with the lowest enthalpy of formation on the map is shown by black hollow circle.

### 3.5. Atomization energy

Figure. 7 shows maps of the atomization energy produced by different MNs. Similar to the enthalpy of formation, lower values of atomization energy are preferred. More negative values of atomization energy (shown in orange and red) means that more energy is required to break all bonds in the crystal. For this property, we took into account the spin-polarization energies of atoms, to take into account that ground states of isolated atoms of most elements are spin-polarized. Among the elements, tungsten has the lowest atomization energy equal to −8.51 eV/atom, while among binary compounds the lowest value is achieved in Ta-C (−8.78 eV/atom for $Ta_6C_5$ and −8.79 eV/atom for $Ta_2C$). Atomization energy measures the total strength of bonding in the solid, and is correlated with the melting temperature. Indeed, tungsten has the highest melting temperature among elements (3695 K), while among binary compounds, HfC and TaC have the highest melting temperatures above 4000 K.[23] For Hf-C, our calculations show atomization energy equal to −8.16 eV/atom. The atomization energy of some representative solids such as graphite, BN (zinc-blende phase), silica ($SiO_2$), and NaCl are −7.98 eV/atom, −7.01 eV/atom, −6.52 eV/atom, and −3.16 eV/atom respectively, which are very close to the values from experiment.

Similar to other properties, i.e. the enthalpy of formation and hardness, AN produces a map with a periodic pattern (Fig.7a), which means clustering materials with similar properties in many small islands instead of few big islands. Looking at the atomization energy maps in the space of MNs in Fig. 7 and their clustering evaluations in Fig.8, and Fig. 9, it is clear that PN and $MN_P$ and $MN_m$ do better job by smoothly clustering materials with similar atomization energy, while clustering rates for USE, and AN are progressively lowered. However, by increasing the $d_p$, number of clusters in AN and USE quickly approaches to the number of clusters in PN, $MN_P$, and $MN_m$, while Fig. 9 shows that number of covered systems by minimum number of clusters ($N_{min}$) for AN is less than all other MNs in all range of $d_p$.

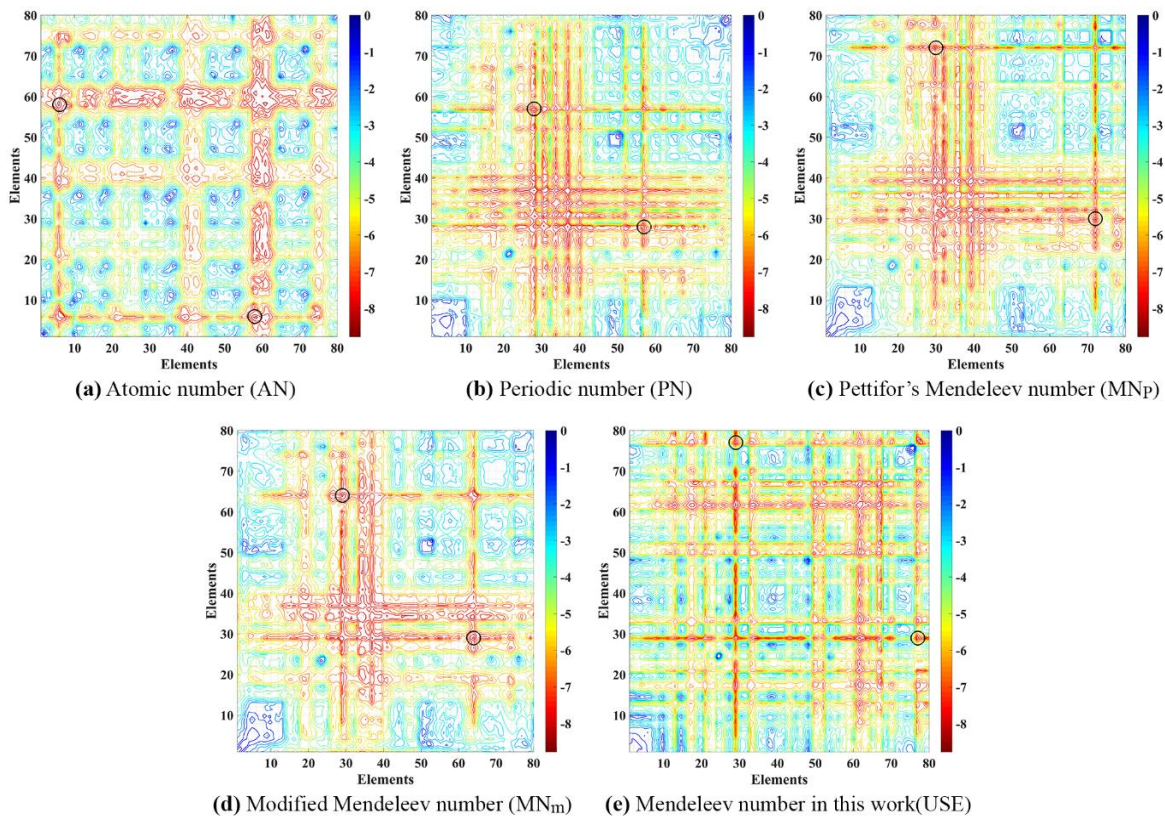

(a) Atomic number (AN)   (b) Periodic number (PN)   (c) Pettifor's Mendeleev number ($MN_P$)

(d) Modified Mendeleev number ($MN_m$)   (e) Mendeleev number in this work(USE)

Figure 7. 2D maps of the atomization energy (eV/atom) of binary systems, plotted in various MNs. The representative for each binary system is a structure with the lowest atomization energy in our database. The material with the lowest atomization energy on the map is shown by black hollow circle.

In a nutshell, except for AN which provides a patchy periodic chemical space, other MNs provide a convenient well-structured chemical space for the properties on which we did tests – hardness (representing the mechanical properties), magnetization (electronic properties), enthalpy of formation and atomization energy (thermochemical properties). Among them, USE, with a simple definition from the most important elemental properties (i.e. atomic radius and electronegativity), generates an overall best clustering in the chemical space (see Table 3., and Fig.8) with clearer separation of regions that contain materials with similar properties. Such well-organized chemical space facilitates the prediction of new materials by exploring the promising regions at the expense of unpromising ones.[22]

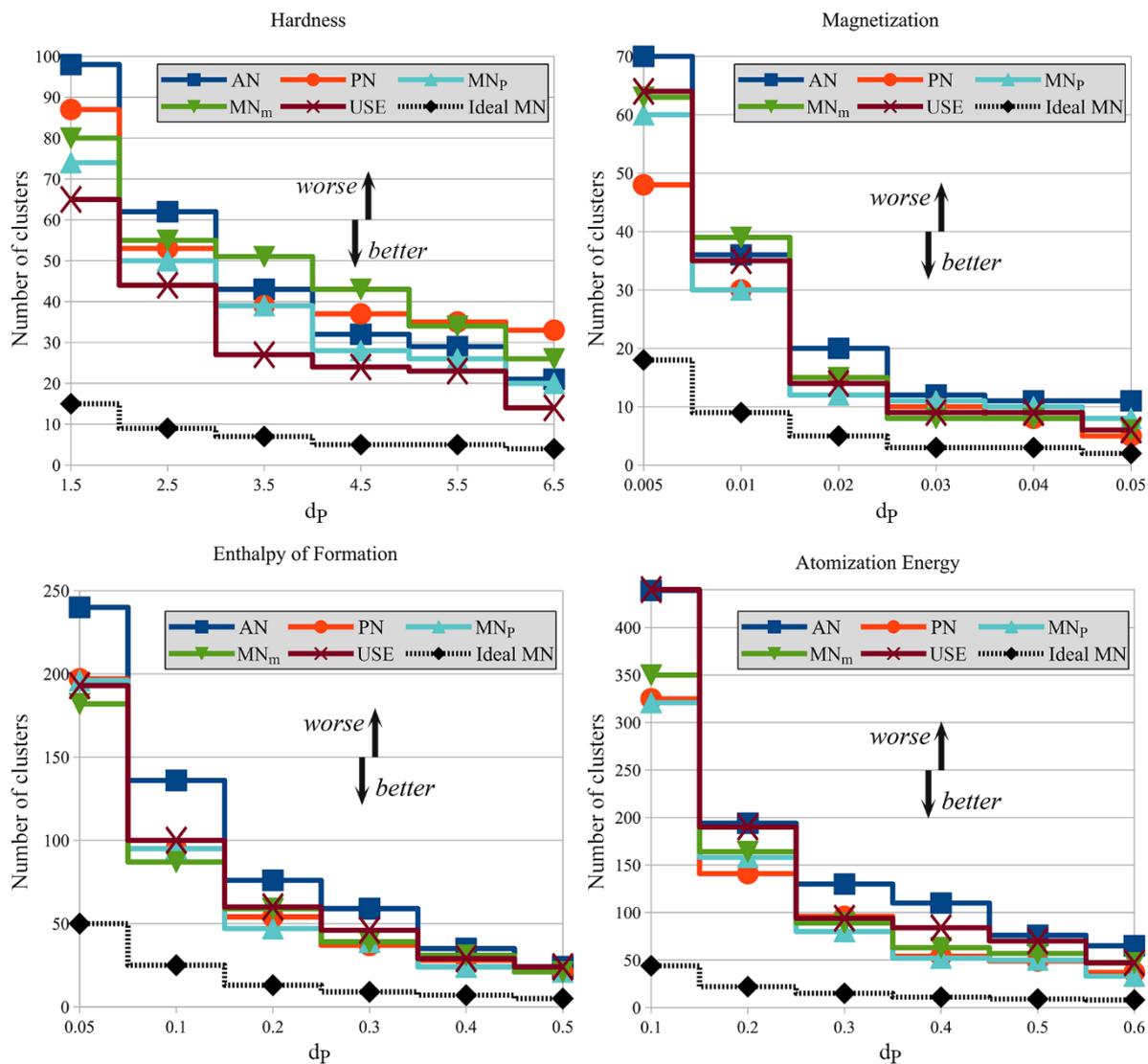

Figure 8. Number of clusters vs. property difference cutoff ($d_P$) for different Mendeleev numbers – in comparison to a hypothetical ideal MN – for the hardness, magnetization, enthalpy of formation, and atomization energy.

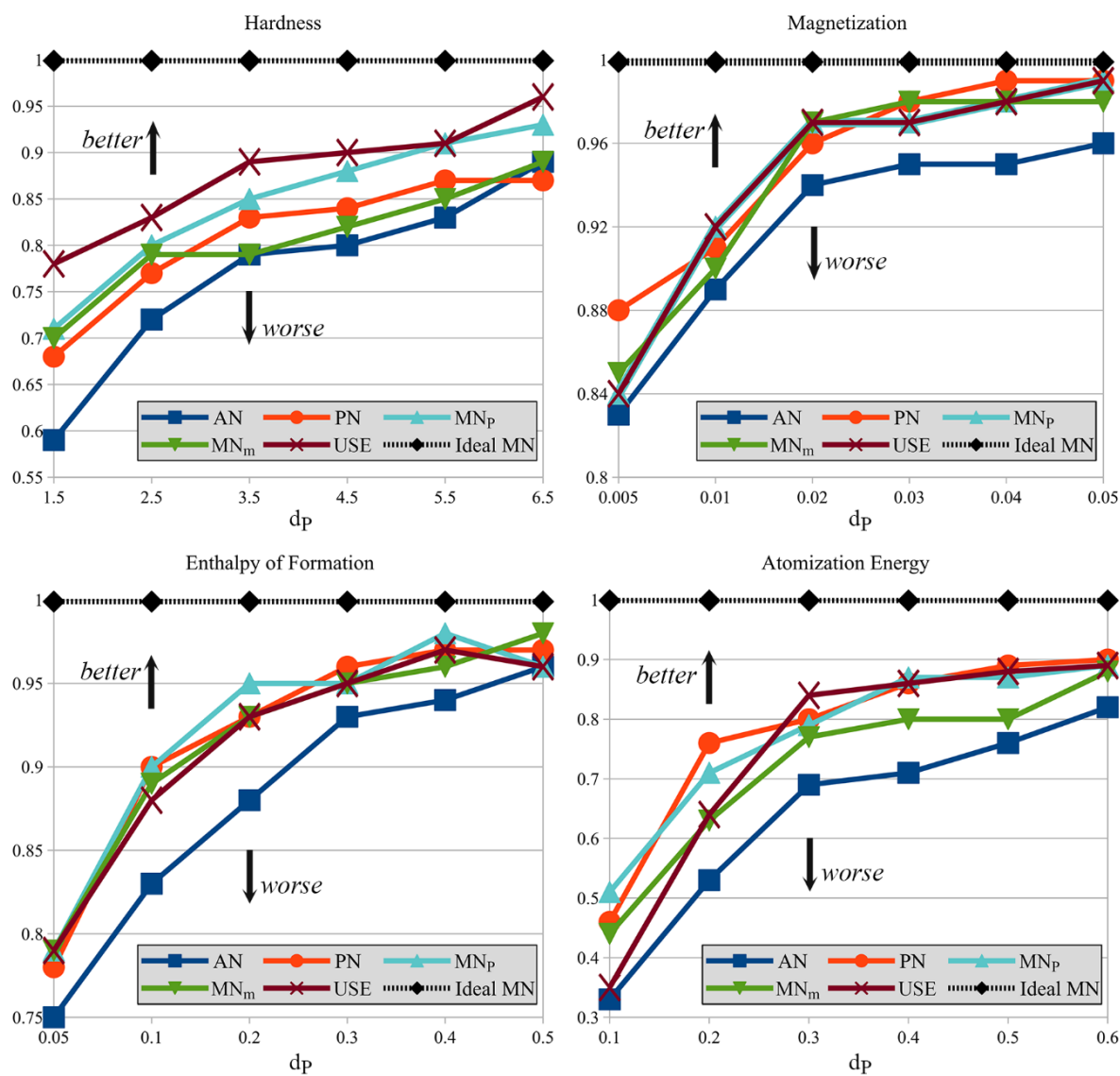

Figure 9. Fraction of binary systems that are covered by a minimum number of clusters as required in an ideal MN to cover all the binary systems for different $d_P$.

### 3.6. A well-defined chemical space at high pressures

The chemistry of the elements and compounds changes with pressure. The discussed MNs are either fixed (AN, PN, and $MN_P$) or obtained by optimizing some evaluation function based on big data ($MN_m$), and adapting these MNs to high pressures is either impossible or requires large amounts of data, huge efforts, and vast computational resources. The USE is the only Mendeleev number that was constructed on a fundamental basis, using the most important elemental properties — electronegativity and atomic radius, and as these properties change under pressure, so will the USE.

The atomic radius of an element can be defined (as we defined throughout this work) and calculated as half the shortest interatomic distance in the relaxed simple cubic structure of that element under pressure. The electronegativity of many elements has been calculated at various pressures.[24,25] Using these data, the USE was obtained at various pressures (Table 4). This can help to predict new materials at arbitrary pressure, only by having a number of relevant data on other systems and plotting them onto the well-organized map produced by the USE.

Table 4. The USE at high pressures.

| # | 50 GPa | 200 GPa | 500 GPa | # | 50 GPa | 200 GPa | 500 GPa |
|---|--------|---------|---------|----|--------|---------|---------|
| 1 | Xe | Cs | Ba | 36 | As | Ir | Ru |
| 2 | Cs | Ba | Cs | 37 | Ge | Nb | Ca |
| 3 | Ba | Po | Bi | 38 | Re | As | As |
| 4 | Po | Bi | Pb | 39 | Ga | Se | Hf |
| 5 | Bi | Pb | Po | 40 | Pt | Pd | Al |
| 6 | Sr | Xe | Sn | 41 | Ti | Sc | Se |
| 7 | Pb | Tl | Tl | 42 | Os | Br | Rh |
| 8 | I | Sn | Xe | 43 | Ir | Ru | Zn |
| 9 | Tl | Sb | Sb | 44 | Tc | Al | Cu |
| 10 | Y | Te | In | 45 | Pd | Na | Sc |
| 11 | Rb | In | Te | 46 | Ru | Ar | Na |
| 12 | Te | I | Hg | 47 | Rh | Rh | Br |
| 13 | Sb | Rb | Rb | 48 | Al | Zn | Cr |
| 14 | Kr | Hg | Cd | 49 | Cl | Ti | Si |
| 15 | Ca | Sr | Au | 50 | V | Cu | Ar |
| 16 | Sn | Y | I | 51 | Zn | V | Nb |
| 17 | In | Cd | Sr | 52 | S | Si | Mn |
| 18 | Lu | Lu | Ag | 53 | Si | Cr | V |
| 19 | Hg | Au | Y | 54 | Cr | Mn | Fe |
| 20 | Hf | Kr | Lu | 55 | P | P | Ni |
| 21 | K | Ta | Zr | 56 | Cu | Fe | Ti |
| 22 | Nb | Ag | W | 57 | Mn | S | P |
| 23 | Sc | Zr | Ta | 58 | Fe | Cl | Co |
| 24 | Cd | W | Re | 59 | Li | Ni | S |
| 25 | Br | Hf | Pt | 60 | Co | Co | Cl |
| 26 | Ta | Re | Os | 61 | Ni | Li | Li |
| 27 | Zr | Pt | Mo | 62 | Ne | Be | Be |
| 28 | Ar | Mo | Kr | 63 | Be | B | B |
| 29 | Na | Os | K | 64 | F | Ne | Ne |
| 30 | Au | K | Ga | 65 | O | C | C |
| 31 | W | Ge | Ge | 66 | B | N | N |
| 32 | S | Ga | Tc | 67 | N | O | O |
| 33 | Mg | Mg | Ir | 68 | C | F | F |
| 34 | Ag | Ca | Pd | 69 | He | He | He |
| 35 | Mo | Tc | Mg | 70 | H | H | H |

## 4. Conclusions

Having a well-defined sequence of the elements (Mendeleev numbers, or MNs), where similar elements take neighboring places, one can produce an organized map of properties for binary or more complex systems that leads to the prediction of new materials by having information on their neighboring systems. We defined a simple, physically meaningful, and universal way to order the elements. In this work, we studied our MN (USE), in addition to a number of previously known MNs such as atomic number (AN), Villars' periodic number [7] (PN), Pettifor's Mendeleev number [2] ($MN_P$), modified Mendeleev number [4] ($MN_m$), using provided data on binary systems from our and other online databases, such as ICSD [8] and COD.[11] Two-dimensional maps of the hardness, magnetization, enthalpy of formation, and atomization energy were plotted using the provided data in the space of MNs and it turned out that most of these sequences (except for AN) indeed work well at clustering materials with similar properties. The evaluation of the MNs showed the overall best clustering rate of the chemical spaces produced by USE for target spaces, i.e. hardness, magnetization, and enthalpy of formation. Also, unlike other MNs, USE can be defined at any arbitrary pressure, which is a step forward for the prediction of materials under pressure. Importantly, our work clarifies the physical meaning of the Mendeleev number (previously defined empirically): it is a collapsed one-number representation of the important atomic properties (such as atomic radius, electronegativity, polarizability, and valence).

**Acknowledgments**

Authors thanks the Russian Science Foundation (grant 19-72-30043) for the financial support.